\documentstyle{article}
\begin{document}
\baselineskip=15pt
\hfuzz=5pt
%----------------------------------------------------------------------
\def\d{\hbox{d}}
\def\e{\hbox{e}}
\def\i{\hbox{i}}
\def\ih{{\i\over\hbar}}
\chardef\ii="10
%----------------------------------------------------------------------
\def\CD{{\cal D}}
\def\CL{{\cal L}}
%----------------------------------------------------------------------
\def\half{{1\over2}}
\def\erfc{\hbox{erfc}}
\def\sign{\hbox{sign}}
\def\SU{\hbox{SU}}
%----------------------------------------------------------------------
\large
\begin{titlepage}
\centerline{July 1993\hfill SISSA/124/93/FM}
\vskip.3in
\begin{center}
{\Large TOWARDS THE CLASSIFICATION OF EXACTLY SOLVABLE FEYNMAN PATH
INTEGRALS: $\delta$-FUNCTION PERTURBATIONS AND BOUNDARY-PROBLEMS AS
MISCELLANEOUS SOLVABLE MODELS}
\vskip.3in
{\Large Christian Grosche$^*$}
\vskip.3in
{\large\em Scuola Internazionale Superiore di Studi Avanzati}
\vskip.1in
{\large\em International School for Advanced Studies}
\vskip.1in
{\large\em Via Beirut 4, 34014 Trieste, Italy}
\end{center}
\normalsize
\baselineskip=12.0pt
\vfill\noindent
Invited talk given at the ``International Workshop on `Symmetry Methods
in Physics' in memory of Ya.\ A.\ Smorodinsky, 5--10 July 1993, Dubna,
Russia; to appear in the proceedings.
\vfill\noindent
\begin{center}
{ABSTRACT}
\end{center}
\noindent
In this contribution I present further results on steps towards a Table
of Feynman Path Integrals. Whereas the usual path integral solutions of
the harmonic oscillator (Gaussian path integrals), of the radial
harmonic oscillator (Besselian path integrals), and the (modified)
P\"oschl-Teller potential(s) (Legendrian path integrals) are well known
and can be performed explicitly by exploiting the convolution properties
of the various types, a perturbative method opens other possibilities
for calculating path integrals. Here I want to demonstrate the
perturbation expansion method for point interactions and boundary
problems in path integrals.

\bigskip\noindent
\centerline{\vrule height0.25pt depth0.25pt width4cm\hfill}
\noindent
{\small $^*$ Address from August 1993: II.Institut f\"ur Theoretische
        Physik, Universit\"at Hamburg, Luruper Chaussee 149,
        22761 Hamburg, Germany.}
\end{titlepage}

\begin{titlepage}
\begin{center}
\ \ \
\end{center}
\end{titlepage}

%-----------------------------------------------------------------------
%                            END OF FILE0
%-----------------------------------------------------------------------

\normalsize
\baselineskip=12.0pt
\noindent{\bf 1.\ Introduction.}
About 25 years ago Smorodinsky, Winternitz et al.\ \cite{MSVW} started
to classify potential problems in quantum mechanics according to their
separability in the Schr\"odinger equation. In particular, they were
interested to find the dynamical symmetry groups of systems in
three-dimensional space, where the Coulomb-potential with the symmetry
group $O(4)$ is just but one, and listed numerous examples of separable
systems, including the separating coordinate systems. It is, of course,
desirable to consider an analogous program in the context of Feynman
path integrals. Whereas no new systems will be found for such a
consideration, the path integral approach is globally in comparison
to the Schr\"odinger approach which is only locally. This opens the
possibility to look at, say, the dynamical symmetries of a system from
a path integral point of view, and to study the system in question by a
group path integration. A comprehensive study along these lines does
not exist yet, but in joint work with F.\ Steiner we are going to
compile a list of exactly solvable Feynman path integrals \cite{GRSf},
a Table of Feynman Path Integrals \cite{GRSg}.

Whereas the usual path integral solutions of the harmonic oscillator
and the general quadratic Lagrangian (Gaussian path integrals), of the
radial harmonic oscillator (Besselian path integrals), and the
(modified) P\"oschl-Teller potential(s) (Legendrian path integrals) are
well known and can be performed explicitly by exploiting the convolution
properties of the various types (i.e.\ they can be seen as a particular
group path integration, respectively a part of it), a perturbative
method opens other possibilities for calculating path integrals and it
is worthwhile to study old problems in this context (e.g.~\cite{LABHb}).
In this contribution I want to demonstrate this approach in a
perturbation expansion for point interactions  and boundary problems in
path integrals. This will include one- , two- and three-dimensional
$\delta$-function perturbations, the one-dimensional $\delta'$-function
perturbation, and Dirichlet and Neumann boundary-conditions,
respectively.

%-----------------------------------------------------------------------
%                            END OF FILE1
%-----------------------------------------------------------------------

\goodbreak
\vglue 0.6truecm
\noindent{\bf 2.\ Formulation of the Path Integral.}
First of all, let us set up the definition of the Feynman path integral.
We first consider the simple case of a classical Lagrangian $\CL(\vec x,
{\dot{\vec x}})=m|{\dot{\vec x}}|^2/2-V(\vec x)$ in $D$ dimensions. Then
the integral kernel ($\vec x\in R^D$)
\begin{equation}
  K(\vec x'',\vec x';t'',t')=
  \Big<\vec x''\Big|\e^{-\i H(t''-t')/\hbar}\Big|\vec x'\Big>
  \Theta(t''-t')\enspace,
\end{equation}
(where $H$ is the Hamiltonian of the system) of the time-evolution
equation
\begin{equation}
  \Psi(\vec x'',t'')=\int_{R^D}
  K(\vec x'',\vec x';t'',t')\Psi(\vec x x',t')d\vec x'\enspace,
\end{equation}
is represented in the form (Feynman path integral \cite{FEYb},
\cite{FH})
\begin{eqnarray}
  K(\vec x'',\vec x';t'',t')
  & =&\lim_{N\to\infty}\bigg({m\over2\pi\i\epsilon\hbar}\bigg)^{ND/2}
   \prod_{j=1}^{N-1}\int_{R^D}dx_{(j)}
  \exp\Bigg\{\ih\sum_{j=1}^N\bigg[{m\over2\epsilon}
  (\vec x_{(j)}-\vec x_{(j-1)})^2-\epsilon V(\vec x_{(j)})\bigg]\Bigg\}
  \nonumber\\
  &\equiv&\int\limits_{\vec x(t')=\vec x'}^{\vec x(t'')=\vec x''}
  \CD\vec x(t)\exp\Bigg\{\ih\int_{t'}^{t''}\bigg[{m\over2}
  |{\dot{\vec x}}|^2-V(\vec x)\bigg]dt\Bigg\}\enspace.
\label{numb}
\end{eqnarray}
Here we have used the abbreviations $\epsilon=(t''-t')/N\equiv T/N$,
$x_{(j)}=x(t_{(j)})$ $(t_{(j)}=t'+\epsilon j,\ j=0,\dots,N)$, and we
interpret the limit $N\to\infty$ as equivalent to $\epsilon\to0$,
$T$ fixed.

The next step is to consider a generic classical Lagrangian of the form
$\CL(\vec q,{\dot{\vec q}}\,)=mg_{ab}(\vec q\,)\dot q^a\dot q^b/2
-V(\vec q\,)$ in some $D$-dimensional Riemannian space $M$ with line
element $ds^2=g_{ab}(\vec q\,)dq^adq^b$. This case, as first
systematically discussed by DeWitt \cite{DEW}, requires a careful
treatment. The Feynman path integral is most conveniently constructed
by considering the Weyl-ordering prescription (e.g.\ \cite{GRSb} and
references therein) in the corresponding quantum Hamiltonian. The
result then is
\begin{eqnarray}
  K(\vec q\,'',\vec q\,';t'',t')
  &=&\big[g(\vec q\,')g(\vec q\,'')\big]^{-1/4}
  \lim_{N\to\infty}\bigg({m\over2\pi\i\epsilon\hbar}\bigg)^{ND/2}
  \prod_{j=1}^{N-1}\int_{M}d\vec q_{(j)}
  \cdot\prod_{j=1}^N\sqrt{g(\bar q_{(j)})}
  \nonumber\\   & &\qquad\times
  \exp\Bigg\{\ih\sum_{j=1}^N\bigg[{m\over2\epsilon}
       g_{ab}(\bar q_{(j)})\Delta q^a_{(j)}\Delta q^b_{(j)}
       -\epsilon V(\bar q_{(j)})
       -\epsilon\Delta V(\bar q_{(j)})\bigg]\Bigg\}\enspace.
\label{numa}
\end{eqnarray}
Here $\bar q_{(j)}=(\vec q_{(j)}+\vec q_{(j-1)})/2$ denotes the midpoint
coordinate, $\Delta\vec q_{(j)}=(\vec q_{(j)}-\vec q_{(j-1)})$, and
$\Delta V(\vec q\,)$ is a well-defined ``quantum potential'' of order
$\hbar^2$ having the form $(\Gamma_a=\partial_a\ln\sqrt{g},\ g=\det
(g_{ab}))$
\begin{equation}
  \Delta V(\vec q\,)={\hbar^2\over8m}
  \Big[g^{ab}\Gamma_a\Gamma_b+2(g^{ab}\Gamma_a)_{,b}
  +{g^{ab}}_{,ab}\Big]\enspace.
\end{equation}
The midpoint prescription together with $\Delta V$ appears in a
completely natural way as an unavoidable consequence of the
Weyl-ordering prescription in the corresponding quantum Hamiltonian
\begin{equation}
  H=-{\hbar^2\over2m}g^{-1/2}\partial_a g^{1/2}g^{ab}\partial_b
    +V(\vec q\,)
  ={1\over8m}\Big(g^{ab}p_ap_b+2p_ag^{ab}p_b+p_ap_bg^{ab}\Big)
     +V(\vec q\,)+\Delta V(\vec q\,)\enspace,
\end{equation}
with $p_a=-\i\hbar(\partial_a+\Gamma_a/2)$, the momentum operator
conjugate to the coordinate $q_a$ in $M$. Of course, choosing another
prescription leads to a different lattice definition in (\ref{numa})
and a different quantum potential $\widetilde{\Delta V}$. However,
every consistent lattice definition of (\ref{numa}) can be transformed
into another one by carefully expanding the relevant metric terms
(integration measure- and kinetic energy term).

Indispensable tools in path integral techniques are transformation
rules. Here coordinate and time transformation must be mentioned which
may be explicitly time-dependent or time-independent. The combination
of these two kind of transformation as developed by Duru and Kleinert
\cite{DK} in the context of the Coulomb problem, boosted enormously
the further development of path integration techniques, for a recent
review see e.g.\ \cite{GROm} for Coulombian problems, and \cite{GRSg}.
However, due lack of space, these transformation techniques will not
be reviewed here, c.f.\ Refs.~\cite{DK}, \cite{FLM}, \cite{GROm},
\cite{GRSb}, \cite{GRSg}, and \cite{KLE}.

%-----------------------------------------------------------------------
%                            END OF FILE2
%-----------------------------------------------------------------------

\goodbreak
\baselineskip=11.5pt
\vglue 0.6truecm
\noindent{\bf 3.\ Basic Path Integrals.}
In this Section we present for completeness some path integrals which
we consider as the Basic Path Integrals, c.f.\ also \cite{GRSf}. From
these path integrals, respectively their Green functions, many other
path integrals can be solved by applying the basic ones.

\goodbreak\bigskip
\noindent{\it 3.1.\ Path Integral for the Harmonic Oscillator.}
The first elementary example is the path integral for the harmonic
oscillator. It has been first evaluated by Feynman \cite{FEYb}.
We have the identity ($\vec x\in R^D$)
\begin{eqnarray}
  & &\int\limits_{\vec x(t')=\vec x'}^{\vec x(t'')=\vec x''}\CD\vec x(t)
  \exp\Bigg[{\i m\over2\hbar}\int_{t'}^{t''}\Big(|{\dot{\vec x}}|^2
       -\omega^2\vec x^2\Big)dt\Bigg]
  \nonumber\\   & &\qquad
  =\sqrt{m\omega\over2\pi\i\hbar\sin\omega T}
   \exp\Bigg\{{\i m\omega\over2\hbar}\bigg[(|\vec x'|^2+|\vec x''|^2)
   \cot\omega T-{2\vec x'\cdot\vec x''\over\sin\omega T}\bigg]\Bigg\}
   \enspace.
\end{eqnarray}
We do not state the expansion into wave-functions ($\propto$ Hermite
polynomials) which can be done by means of the Mehler formula, nor the
corresponding Green's function.
\newline
The path integral for quadratic Lagrangians can also be stated exactly
\begin{equation}
  \int\limits_{\vec x(t')=\vec x'}^{\vec x(t'')=\vec x''}\CD\vec x(t)
  \exp\bigg(\ih\int_{t'}^{t''}\CL(\vec x,{\dot{\vec x}})dt\bigg)
  =\bigg({1\over2\pi\i\hbar}\bigg)^{D/2}
  \sqrt{\det\bigg(-{\partial^2 S_{Cl}[\vec x'',\vec x']\over
        \partial x_a''\partial x_b'}\bigg)}
  \exp\bigg(\ih S_{Cl}[\vec x'',\vec x']\bigg)\enspace.
\label{numj}
\end{equation}
Here $\CL(\vec x,{\dot{\vec x}})$ denotes any classical Lagrangian at
most quadratic in $\vec x$ and ${\dot{\vec x}}$, and $S_{Cl}[\vec x'',
\vec x']=\int_{t'}^{t''}dt$ $\CL(\vec x_{Cl},{\dot{\vec x}}_{Cl})$ the
corresponding classical action evaluated along the classical solution
$\vec x_{Cl}$ satisfying the boundary conditions $\vec x_{Cl}(t')
=\vec x'$, $\vec x_{Cl}(t'')=\vec x''$. The determinant appearing in
(\ref{numj}) is known as the van Vleck-Pauli-Morette determinant (see
e.g.\ \cite{DEW} and references therein). The explicit evaluation of
$S_{Cl}[\vec x'',\vec x']$ may have any degree of complexity due to
complicated classical solutions of the Euler-Lagrange equations as the
classical equations of motion. Of course, the harmonic oscillator is
but a simple example of (\ref{numj}).

\goodbreak\bigskip
\noindent{\it 3.2.\ Path Integral for the Radial Harmonic Oscillator.}
In order to evaluate the path integral for the radial harmonic
oscillator, one has to perform a separation of the angular variables,
see Refs.~\cite{GOOb}, \cite{PI}. Path integrals related to the radial
harmonic oscillator may be called of Besselian type \cite{INO}. Here we
are not going into the subtleties of the Besselian functional measure
due to the Bessel functions which appear in the lattice approach
\cite{FLM}, \cite{GOOb}, \cite{GRSb}, \cite{PI}, \cite{STEc} which is
actually necessary for the explicit evaluation of the radial harmonic
oscillator path integral. One obtains (modulo the above mentioned
subtleties) ($r>0$)
\begin{eqnarray}
  & &\int\limits_{r(t')=r'}^{r(t'')=r''}\CD r(t)
  \exp\bigg[\ih\int_{t'}^{t''}\bigg({m\over2}\dot r^2
  -\hbar^2{\lambda^2-{1\over4}\over2mr^2}
           -{m\over2}\omega^2r^2\bigg)dt\bigg]
  \nonumber\\   & &\qquad
  =\sqrt{r'r''}{m\omega\over \i\hbar\sin\omega T}
  \exp\bigg[-{m\omega\over2\i\hbar}({r'}^2+{r''}^2)\cot\omega T\bigg]
  I_\lambda\bigg({m\omega r'r''\over \i\hbar\sin\omega T}\bigg)\enspace,
\end{eqnarray}
where $I_\lambda(z)$ denotes the modified Bessel function.

\eject
\baselineskip=12.0pt
\noindent{\it 3.3.\ Path Integral for the P\"oschl-Teller Potential.}
There are two further basic path integral solutions based on the
$\SU(2)$ \cite{BJb}, \cite{INOWI} and $\SU(1,1)$ \cite{BJb} group path
integration, respectively. These path integrations may be called of
Legendrian type \cite{INO}. The first yields the path integral identity
for the solution of the P\"oschl-Teller potential according to
($0<x<\pi/2$)
\begin{eqnarray}
  & &\ih\int_0^\infty dT \e^{\i ET/\hbar}
  \int\limits_{x(t')=x'}^{x(t'')=x''}\CD x(t)
  \exp\Bigg\{\ih\int_{t'}^{t''}\bigg[{m\over2}\dot x^2
        -{\hbar^2\over2m}\bigg({\kappa^2-{1\over4}\over\sin^2x}
   +{\lambda^2-{1\over4}\over\cos^2x}\bigg)\bigg]dt\Bigg\}
  \nonumber\\   & &\quad
  ={m\over\hbar^2}\sqrt{\sin2x'\sin2x''}
  {\Gamma(m_1-L_E)\Gamma(L_E+m_1+1)\over
   \Gamma(m_1+m_2+1)\Gamma(m_1-m_2+1)}
  \nonumber\\   & &\qquad\times
  \bigg({1-\cos2x_<\over2}\bigg)^{(m_1-m_2)/2}
  \bigg({1+\cos2x_<\over2}\bigg)^{(m_1+m_2)/2}
  \nonumber\\   & &\qquad\times
  \bigg({1-\cos2x_>\over2}\bigg)^{(m_1-m_2)/2}
  \bigg({1+\cos2x_>\over2}\bigg)^{(m_1+m_2)/2}
  \nonumber\\   & &\qquad\times
  {_2}F_1\bigg(-L_E+m_1,L_E+m_1+1;m_1-m_2+1;{1-\cos2x_<\over2}\bigg)
  \nonumber\\   & &\qquad\times
  {_2}F_1\bigg(-L_E+m_1,L_E+m_1+1;m_1+m_2+1;{1+\cos2x_>\over2}\bigg)
\end{eqnarray}
with $m_{1/2}=\half(\lambda\pm\kappa)$, $L_E=-\half+\half\sqrt{2mE}\,
/\hbar$, and $x_{<,>}$ the larger, smaller of $x',x''$, respectively.
${_2}F_1(a,b;c;z)$ denotes the hypergeometric function. Here we have
used the fact that it is possible to state closed expressions for the
(energy dependent) Green's functions for the P\"oschl-Teller (and
modified P\"oschl-Teller potential, respectively), by summing up the
spectral expansion, c.f.\ \cite{KLEMUS}. The case of the modified
P\"oschl-Teller potential will not be given here. It follows from a
properly chosen coordinate transformation of the P\"oschl-Teller
potential case \cite{KLEMUS}.

%-----------------------------------------------------------------------
%                            END OF FILE3
%-----------------------------------------------------------------------

\goodbreak
\vglue 0.6truecm
\noindent{\bf 4.\ Point Interactions.}
The general method for the time-ordered perturbation expansion is
simple. We assume that we have a potential $W(\vec x)=V(\vec x)+
\tilde V(\vec x)$ ($\vec x\in R^D$) in the path integral, where it is
assumed that $W$ is so complicated that a  direct path integration is
not possible. However, the path integral corresponding to $V(\vec x)$
is assumed to be known, which we call $K^{(V)}(T)$. We expand the path
integral containing $V(\vec x)$ in a perturbation expansion about
$\tilde V(\vec x)$ in the following way. The initial kernel
corresponding to $V$ propagates in $\epsilon$-time unperturbed, then it
is interacting with $\tilde V$, propagates again in another
$\epsilon$-time unperturbed, a.s.o, up to the final state. Let us
denote the path integral corresponding to the potential $V$ by
\begin{equation}
  K^{(V)}(\vec x'',\vec x';T)
  =\int\limits_{\vec x(t')=\vec x'}^{\vec x(t'')=\vec x''}
  \CD\vec x(t)\exp\Bigg\{\ih\int_{t'}^{t''}
  \bigg[{m\over2}|{\dot{\vec x}}|^2-V(\vec x)\bigg]dt\Bigg\}\enspace.
\end{equation}
We introduce the (energy-dependent) Green function (resolvent kernel)
\begin{eqnarray}
  G^{(V)}(\vec x'',\vec x';E)
  &=&\ih\int_0^\infty dT\,\e^{\i ET/\hbar}K^{(V)}(\vec x'',\vec x';T)
  \\
  K^{(V)}(\vec x'',\vec x';T)
  &=&{1\over2\pi\i}\int_{-\infty}^{\infty}
  G^{(V)}(\vec x'',\vec x';E)\e^{-\i ET\hbar}dE\enspace.
\end{eqnarray}
This gives the series expansion (see also e.g.\ \cite{BAU}, \cite{FH},
\cite{GODE}, \cite{GROh}, \cite{LABH}, \cite{SCHUe})
\begin{eqnarray}
  & &K(\vec x'',\vec x';T)
  =\int\limits_{\vec x(t')=\vec x'}^{\vec x(t'')=\vec x''}\CD\vec x(t)
  \exp\Bigg\{\ih\int_{t'}^{t''}\bigg[
  {m\over2}|{\dot{\vec x}}|^2-V(\vec x)-\tilde V(\vec x)\bigg]dt\Bigg\}
  \nonumber\\   & &
  =K^{(V)}(\vec x'',\vec x';T)+\sum_{n=1}^\infty\bigg(-\ih\bigg)^n
  \prod_{j=1}^n\int_{t'}^{t_{j+1}} dt_j\int_{R^D}d\vec x_j
  \nonumber\\   & &\qquad\times
  \vphantom{\bigg]^{1/2}}
  K^{(V)}(\vec x_1,\vec x';t_1-t')
  \widetilde V(x_1)K^{(V)}(\vec x_2,\vec x_1;t_2-t_1)
  \times\dots
  \nonumber\\   & &\qquad\dots\times
  \widetilde V(x_{n-1})K^{(V)}(\vec x_n,\vec x_{n-1};t_n-t_{n-1})
  \widetilde V(\vec x_n)K^{(V)}(\vec x'',\vec x_n;t''-t_n)\enspace.
  \vphantom{\bigg]^{1/2}}
\end{eqnarray}
Here we have ordered the time as $t'=t_0<t_1<t_2<\dots<
t_{n+1}=t''$ and paid attention to the fact that $K^{(V)}(t_j-t_{j-1})$
is different from zero only if $t_j>t_{j-1}$.

\goodbreak\bigskip
\noindent{\it 4.1.\ $\delta$-Function Perturbations in One Dimension.}
Let us consider first $D=1$ and $\widetilde V(x)=-\gamma\delta(x-a)$.
Introducing the Green function $G^{(\delta)}(E)$ for the perturbed
problem similarly as for $G^{(V)}(E)$, it is due to the convolution
theorem possible to sum up the perturbation series and one obtains
\begin{eqnarray}
  & &\ih \int_0^\infty  dT\,\e^{\i TE/\hbar}
   \int\limits_{x(t')=x'}^{x(t'')=x''}\CD x(t)
  \exp\Bigg\{\ih\int_{t'}^{t''}
  \bigg[{m\over2}\dot x^2-V(x)+\gamma\delta(x-a)\bigg]dt\Bigg\}
  \nonumber\\   & &\qquad
  =G^{(V)}(x'',x';E)+
   \frac{G^{(V)}(x'',a;E)G^{(V)}(a,x';E)}
        {1/\gamma-G^{(V)}(a,a;E)}\enspace.
\label{numc}
\end{eqnarray}
An implicit equation for the propagator has been stated by Gaveau and
Schulman \cite{GASCH}. The simple example for a $\delta$-function
perturbation with $V\equiv0$ has been discussed by several authors,
e.g.\ \cite{BAU}, \cite{GODE}, \cite{GROh}, \cite{LABH}, and it is
possible to state explicitly the corresponding propagator
$K^{(\delta)}(T)$. In a similar way, one treats the case of a radial
$\delta$-function perturbation, so called shell-interactions
\cite{AGHHb}, \cite{GROh} ($\vec x\in R^D$):
\begin{eqnarray}
  & &\ih \int_0^\infty  dT\,\e^{\i TE/\hbar}
  \int\limits_{\vec x(t')=\vec x'}^{\vec x(t'')=\vec x''}\CD\vec x(t)
  \exp\Bigg\{\ih\int_{t'}^{t''}
 \bigg[{m\over2}|{\dot{\vec x}}|^2-V(r)+\gamma\delta(r-a)\bigg]dt\Bigg\}
  \nonumber\\   & &\qquad
  =\sum_{l=0}^\infty  S_l^\mu(\Omega'')S_l^{\mu*}(\Omega')
  \bigg[G_l^{(V)}(r'',r';E)
  +\frac{G_l^{(V)}(r'',a;E)G_l^{(V)}(a,r';E)}
  {a^{1-D}/\gamma-G_l^{(V)}(a,a;E)}\bigg]\enspace.
\end{eqnarray}
Here the $S_l^\mu(\Omega)$ denote the real hyperspherical harmonics on
the sphere $S^{(D-1)}$. It is obvious that more than one
$\delta$-function perturbation can be taken into account, in one
dimension as well as in the radial case, c.f.\ \cite{AGHHb},
\cite{GROx}, \cite{GROw} for details. In the limit of infinitely many
singular perturbations a lattice is  obtained, c.f.\ \cite{AGHHb} and
\cite{GOBR}.

\goodbreak\bigskip
\noindent{\it 4.2.\ $\delta'$-Function Perturbations in One Dimension.}
Due to the specific property of one-dimensional space, it is also
possible to study $\delta'$-interactions in one dimension. The
$\delta'$-function perturbation, also called dipole interaction, is a
rather strange object and there seems to no way to generate this kind
of interaction form the usual $\delta$-function perturbations, e.g.\ by
an appropriate limiting procedure in the definition of the derivative.
However, if one makes the delay over the one-dimensional Dirac equation
and its corresponding path integral representation \cite{FH} it is
possible to incorporate $\delta'$ interactions \cite{GROz}: One
considers a $\delta$-function perturbation in the path integral
representation for the one-dimensional Dirac equation. If one chooses
the ``electron'' component, one recovers in the non-relativistic limit
the usual $\delta$-function perturbation. If one chooses the
``positron'' component, a $\delta'$-function perturbation is generated
in the non-relativistic limit \cite{AGHHb}. The result then has the form
\begin{eqnarray}
  & &\ih                    \int_0^\infty dT\,\e^{\i ET/\hbar}
  \int\limits_{x(t')=x'}^{x(t'')=x''}\CD x(t)
  \exp\Bigg\{\ih\int_{t'}^{t''}\bigg[{m\over2}\dot x^2
       -V(x)\bigg]dt\Bigg\}
  \nonumber\\   & &\qquad
  =G^{(V)}(x'',x';E)
   -\frac{G^{(V)}_{,x'}(x'',a;E)G^{(V)}_{,x''}(a,x';E)}
   {1/\gamma+{\widehat G}^{(V)}_{,x'x''}(a, a;E)}\enspace,
\label{numd}
\end{eqnarray}
where ${\widehat G}^{(V)}_{,x'x''}(a, a;E)$ denotes a regularized Green
function with all infinities subtracted. For the e.g.\ $V\equiv0$ case
one obtains
\begin{eqnarray}
  & &\ih\int_0^\infty dT\,\e^{\i ET/\hbar}
  \int\limits_{x(t')=x'}^{x(t'')=x''}\CD x(t)
  \exp\Bigg\{\ih\int_{t'}^{t''}
  \bigg[{m\over2}\dot x^2+\gamma\delta'(x-a)\bigg]dt\Bigg\}
  \nonumber\\   & &\quad
  ={1\over\hbar}\sqrt{-{m\over2E}}
   \exp\bigg(-{\sqrt{-2mE}\over\hbar}|x''-x'|\bigg)
  \nonumber\\   & &\qquad
  -{m^2\over\hbar^4}{\exp\Big[-\sqrt{-2mE}\,
          \big(|x''-a|+|a-x'|\big)/\hbar\Big]
          \over1/\gamma-m\sqrt{-2mE}/\hbar^3}
  \,\sign(x''-a)\sign(x'-a)\enspace.
\end{eqnarray}
\baselineskip=12.5pt
The corresponding propagator is given by
\begin{eqnarray}
  & &\int\limits_{x(t')=x'}^{x(t'')=x''}\CD x(t)
  \exp\Bigg\{\ih\int_{t'}^{t''}
  \bigg[{m\over2}\dot x^2+\gamma\delta'(x-a)\bigg]dt\Bigg\}
  \nonumber\\   & &\quad
=\sqrt{m\over2\pi\i\hbar T}\exp\bigg[-{m\over2\i\hbar T}(x''-x')^2\bigg]
  \nonumber\\   & &\qquad
  +\sqrt{m\over2\pi\i\hbar T}
   \exp\bigg[-{m\over2\i\hbar T}(|x''-a|+|x'-a|)^2\bigg]
  \,\sign(x''-a)\sign(x'-a)
  \nonumber\\   & &\qquad
  +{\hbar^2\over2m\gamma}\exp\Bigg[-{\hbar^2\over m\gamma}
        \Big(|x''-a|+|x'-a|\Big)+\ih{\hbar^6T\over2m^3\gamma^2}\Bigg]
  \nonumber\\   & &\qquad\quad\times
  \erfc\Bigg\{\sqrt{m\over2\i\hbar T}\,\bigg[\Big(|x''-a|+|x'-a|\Big)
       -{\i\hbar^3 T\over m^2\gamma}\bigg]\Bigg\}\,
  \sign(x''-a)\sign(x'-a)\enspace.
\end{eqnarray}
For $\gamma>0$ there is one bound state with energy level and
wave-function given by
\begin{equation}
  E^{(\delta')}=-{\hbar^6\over2m^3\gamma^2},\qquad
  \Psi^{(\delta')}(x)={\hbar\over\sqrt{m\gamma}}
  \exp\bigg(-{\hbar^2\over m\gamma}|x-a|\bigg)\,\sign(x-a)\enspace.
\end{equation}

\goodbreak\bigskip
\noindent{\it 4.3.\ $\delta$-Function Perturbations in Two and Three
Dimensions.}
We consider $\delta$-function perturbations in two and three dimensions.
A na\"\ii ve generalization of the corresponding one-dimensional result
gives a divergence. The origin of this divergence can be easily seen.
One must evaluate the corresponding Green function of the unperturbed
problem at both arguments being equal, an expression which diverges
logarithmically in two dimensions, and in three dimensions there is a
simple pole. Consequently, one must regularize the problem which
basically consists of finding the proper Friedrich extension in order
to make the corresponding Hamiltonian self-adjoint. This topic is
addressed in a comprehensive way by Albeverio et al.\ \cite{AGHHb} and
has been put into the path integral language in \cite{GROy}. The main
idea is that in the regularization procedure a new (regularized)
coupling $\alpha$ is introduced, while letting $\gamma$ in the
heuristic expression ``$\gamma\delta(\vec x-\vec a)$'' be zero in a
suitable way, as first pointed out be Berezin and Faddeev \cite{BF}. We
just cite the result: For the Green function of a $\delta$-function
perturbation in two and three dimensions one obtains:
\begin{eqnarray}\bigskip
  & &\ih\int_0^\infty dT\,\e^{\i ET/\hbar}
  \int\limits_{\vec x(t')=\vec x'}^{\vec x(t'')=\vec x''}
  \CD_{\Gamma^{(V)}_{\alpha,\vec a}}\vec x(t)
  \exp\Bigg\{\ih\int_{t'}^{t''}\bigg[{m\over2}|{\dot{\vec x}}|^2
  -V(\vec x)+\gamma\delta(\vec x-\vec a)\bigg]dt\Bigg\}
  \nonumber\\   & &\qquad
 =G^{(V)}(\vec x'',\vec x';E)+\big(\Gamma_{\gamma,\vec a}^{(V)}\big)^{-1}
  G^{(V)}(\vec x'',\vec a;E)G^{(V)}(\vec a,\vec x';E)\enspace.
\end{eqnarray}
with $\Gamma_{\gamma,\vec a}^{(V)}$ given by $\Gamma_{\gamma,\vec a}
^{(V)}=\alpha g_{0,\lambda}-g_{1,\lambda}$. Here denote ($\alpha\in
(-\infty,\infty]$, $r=|\vec x|$)
\begin{equation}
  g_{0,\lambda}=\lim_{r\to0}{g(r)\over G_\lambda^{(0)}(r)}
  \enspace,\qquad
  g_{1,\lambda}=\lim_{r\to0}
  {[g(r)-g_{0,\lambda}G_\lambda^{B}(r)]\over F_\lambda^{(0)}(r)}
  \enspace,
\end{equation}
where $F_\lambda^{(0)}(r)=r^\lambda$, $G^{(0)}_\lambda(r)=-(m/\pi
\hbar^2) \sqrt{r}\,\ln r$ for $D=2$, $G^{(0)}_\lambda(r)=-(m/2\pi
\hbar^2)$ for $D=3$; $G^B_\lambda(r)$ denotes the asymptotic expansion
of the irregular solution of the corresponding Schr\"odinger problem up
to order $r^t$, $t\leq2\lambda-1$; a convenient way for choosing $g(r)$
is to take the reduced Green function of the unperturbed problem, i.e.\
$g(r)=\Omega^{-1}(D) r^\lambda G^{(V)}(r,r;E)$, with $\Omega(D)$ the
volume of the $D$-dimensional unit-sphere  (for more details, c.f.\
\cite{AGHHb} and \cite{GROy} for a path integral discussion including
some instructive examples). In our notation the index $\Gamma_{\gamma,
\vec a}^{(V)}$ in $\CD_{\Gamma_{\gamma,\vec a}^{(V)}}(x)$ denotes the
to-be-performed regularization in the path integral. The connection
between $\lambda$ and the dimension $D$ is given by $\lambda=(D-1)/2$.
For a three-dimensional singular perturbation with $V\equiv0$ it is
possible to state the corresponding propagator in a straightforward
way, c.f.\ \cite{SCATE}, whereas in two dimensions only a rather
complicated integral representation can be given, c.f.\ \cite{ABD} and
\cite{GROy}.

%-----------------------------------------------------------------------
%                            END OF FILE4
%-----------------------------------------------------------------------

\goodbreak
\vglue 0.6truecm
\noindent{\bf 5.\ Boundary Conditions.} The results from the
incorporation of the $\delta$- and $\delta'$-function perturbation in
one dimension make it possible to build in Dirichlet and Neumann
boundary-conditions, respectively, in the path integral \cite{GROw}.
Some general, however not explicit, results have been archived in
Refs.~\cite{CAR} and \cite{CFG}, and a particular example (the free
particle) was discussed in \cite{CMS}. Making now in our formalism the
strength of the $\delta$-function perturbations in (\ref{numc})
infinitely repulsive produces Dirichlet boundary-conditions (D) at the
location of the $\delta$-function, i.e., we obtain a path integral
representation in a half-space with a boundary with Dirichlet
boundary-conditions at $x=a$:
\begin{eqnarray}
  & &\ih                    \int_0^\infty dT\,\e^{\i ET/\hbar}
  \int\limits_{x(t')=x'}^{x(t'')=x''}\CD_{(Wall)}^{(D)}x(t)
  \exp\Bigg\{\ih\int_{t'}^{t''}\bigg[{m\over2}\dot x^2
       -V(x)\bigg]dt\Bigg\}
  \nonumber\\   & &\qquad
  =G^{(V)}(x'',x';E)
   -\frac{G^{(V)}(x'',a;E)G^{(V)}(a,x';E)}{G^{(V)}(a,a;E)}\enspace.
\end{eqnarray}
Note that for a symmetrical model, i.e.\ $V(x)=V(-x)$, a
$\delta$-function perturbation at $x=0$ yields in the limit
$\gamma\to\infty$ a doubly generated energy level spectrum with one
``spurious ground state'' $\tilde\Psi_0$ with infinite negative energy
and a corresponding wave-function concentrated at $x=0$, i.e.\ $\vert
\tilde\Psi_0(x)\vert^2=\delta(x)$ (e.g.~\cite{AGSTA} and references
therein).

Similarly, making the strength of the $\delta'$-function perturbations
in (\ref{numd}) infinitely repulsive produces Neumann
boundary-conditions (N) at the location of the $\delta'$-function,
i.e., we obtain a path integral representation in half-space with a
boundary with Neumann boundary-conditions at $x=a$ \cite{GROz}:
\begin{eqnarray}
  & &\ih                    \int_0^\infty dT\,\e^{\i ET/\hbar}
  \int\limits_{x(t')=x'}^{x(t'')=x''}\CD_{(Wall)}^{(N)}x(t)
  \exp\Bigg\{\ih\int_{t'}^{t''}\bigg[{m\over2}\dot x^2
       -V(x)\bigg]dt\Bigg\}
  \nonumber\\   & &\qquad
  =G^{(V)}(x'',x';E)
   -\frac{G^{(V)}_{,x'}(x'',a;E)G^{(V)}_{,x''}(a,x';E)}
   {\widehat{G}^{(V)}_{,x'x''}(a, a;E)}\enspace.
\end{eqnarray}

We can further consider motion in boxes, where any combination of
Dirichlet and Neumann boundary-conditions is allowed, e.g.: For the
motion in the box $a<x<b$ with Neumann boundary-conditions  for $x=a$
and $x=b$ we obtain:
\begin{eqnarray}
  & & \ih                    \int_0^\infty dT\,\e^{\i ET/\hbar}
  \int\limits_{x(t')=x'}^{x(t'')=x''}\CD_{(a<x<b)}^{(NN)}x(t)
  \exp\Bigg\{\ih\int_{t'}^{t''}\bigg[{m\over2}\dot x^2
       -V(x)\bigg]dt\Bigg\}
  \nonumber\\   & &\qquad
  =\frac{\left|\begin{array}{ccc}
          G^{(V)}(x'',x';E) &G^{(V)}_{,x'}(x'',b;E)
                        &G^{(V)}_{,x'}(x'',a;E)       \\
          G^{(V)}_{,x''}(b,x';E)
                        &\widehat{G}^{(V)}_{,x'x''}(b,b;E)
                        &\widehat{G}^{(V)}_{,x'x''}(b,a;E)\\
          G^{(V)}_{,x''}(a,x';E)
                        &\widehat{G}^{(V)}_{,x'x''}(a,b;E)
                        &\widehat{G}^{(V)}_{,x'x''}(a,a;E)
  \end{array}\right|}{\left|\begin{array}{cc}
  \widehat{G}^{(V)}_{,x'x''}(b,b;E)
                        &\widehat{G}^{(V)}_{,x'x''}(b,a;E)    \\
  \widehat{G}^{(V)}_{,x'x''}(a,b;E)
                        &\widehat{G}^{(V)}_{,x'x''}(a,a;E)
  \end{array}\right|}\enspace.
\end{eqnarray}

Finally, we can put the results of the motion in a half-space with
Dirichlet {\it and\/} Neumann boundary-conditions together to find a
path integral representation for potential problems with absolute value
dependence:
\begin{eqnarray}
  & &\ih                    \int_0^\infty dT\,\e^{\i ET/\hbar}
   \int\limits_{x(t')=x'}^{x(t'')=x''}\CD x(t)
  \exp\Bigg\{\ih\int_{t'}^{t''}\bigg[{m\over2}\dot x^2
       -V(|x|)\bigg]dt\Bigg\}
  \nonumber\\   & &\qquad
  =G^{(V)}(x'',x';E)
   -\frac{G^{(V)}(x'',0;E)G^{(V)}(0,x';E)}{2G^{(V)}(0,0;E)}
   -\frac{G^{(V)}_{,x'}(x'',0;E)G^{(V)}_{,x''}(0,x';E)}
   {2\widehat{G}^{(V)}_{,x'x''}(0,0;E)}\enspace.
\end{eqnarray}
Important applications of the latter equation are the so-called
``double-oscillator'' $V^{(DO)}(x)$ $ =(m/2)$ $\omega^2$ $(|x|-a)^2$
and the one-dimensional Kepler problem $V^{(1-D)}(x)=k|x|$.

%-----------------------------------------------------------------------
%                            END OF FILE5
%-----------------------------------------------------------------------

\goodbreak
\vglue 0.6truecm
\noindent{\bf 6.\ Summary.}
In this contribution I have presented some recent developments in path
integral techniques, i.e.\ the incorporation of point interactions and
boundary problems in the path integral. They were $\delta$-function
perturbations in one dimension, $\delta'$-function perturbations in
one dimension, and $\delta$-function perturbations in two and three
dimensions. The results for the $\delta$- and $\delta'$-function
perturbations, respectively, made it possible to incorporate Dirichlet
and Neumann boundary-conditions in the path integral. This could be
archived by making the strength of the singular perturbations infinitely
repulsive. All the corresponding Green functions were stated, including
some important combinations, for the quantum motion in boxes with
Dirichlet and/or Neumann boundary conditions at the walls, and for
potential problems with absolute value dependence.

\goodbreak
\vglue 0.6truecm
\noindent{\bf Acknowledgements.}
I want to thank the organizers of the Dubna workshop for the nice
atmosphere and warm hospitality. In particular I want to thank V.\ V.\
Belokurov, L.\ S.\ Davtian, A.\ Inomata, G.\ Junker, R.\ M.\
Mir-Kasimov, G.\ S.\ Pogosyan, O.\ G.\ Smolyanov, and S.\ I.\ Vinitsky
for fruitful discussions. Furthermore I want to thank F.\ Steiner
(Hamburg University) for the joint work on a Table of Feynman Path
Integrals from which this contribution is a part of.

%-----------------------------------------------------------------------
%                            END OF FILE6
%-----------------------------------------------------------------------

\end{document}